\documentclass[aps,PRL,twocolumn,groupedaddress,floats]{revtex4}
\usepackage{amssymb}
\usepackage{graphicx}

\begin{document}

\title{Correlation effects of exchange splitting and coexistence of spin-density-wave and superconductivity in single crystalline Sr$_{1-x}$K$_x$Fe$_2$As$_2$ }

\author{Y. Zhang$^{1}$,  J. Wei$^{1}$, H. W. Ou$^{1}$,  J. F. Zhao$^{1}$, B. Zhou$^{1}$,
F. Chen$^{1}$, M. Xu$^{1}$, C. He$^{1}$, G. Wu$^2$, H. Chen$^2$, M.
Arita$^{3}$, K. Shimada$^{3}$, H. Namatame$^{3}$, M.
Taniguchi$^{3}$, X. H. Chen$^2$, D. L. Feng$^{1}$}
\email{dlfeng@fudan.edu.cn}

\affiliation{$^1$Department of Physics, Surface Physics Laboratory
(National Key Laboratory), and Advanced Materials Laboratory, Fudan
University, Shanghai 200433, P. R. China}

\affiliation{$^2$Hefei National Laboratory for Physical Sciences at
Microscale and Department of Physics, University of Science and
Technology of China, Hefei, Anhui 230026, P. R. China}

\affiliation{$^3$Hiroshima Synchrotron Radiation Center and Graduate
School of Science, Hiroshima University, Hiroshima 739-8526, Japan.}

\date{\today}

\begin{abstract}

The nature of spin-density wave and its relation with
superconductivity are crucial issues in the newly discovered
Fe-based high temperature superconductors. Particularly it is
unclear whether the superconducting phase and spin density wave
(SDW) are truly exclusive from each other as suggested by certain
experiments. With angle resolved photoemission spectroscopy, we here
report exchange splittings of the band structures in
Sr$_{1-x}$K$_{x}$Fe$_2$As$_2$ ($x=0,\,0.1,\,0.2$), and the
non-rigid-band behaviors of the splitting. Our data on single
crystalline superconducting samples unambiguously prove that SDW and
superconductivity could coexist in iron-pnictides.

\end{abstract}


\maketitle


Both the cuprates and the iron pnictides high temperature
superconductors are in the vicinity of certain magnetic order
\cite{PCDai}. For the cuprate, the antiferromagnetic spin
fluctuations might likely facilitate the $d$-wave pairing, which
makes the nature of the spin density wave (SDW) in the iron
pnictides and its relation with the superconductivity central
issues. Recently, we found that exchange splittings of the bands
(instead of Fermi surface nesting) are responsible for the SDW
formation in BaFe$_2$As$_2$\cite{YangExchange}. This is beyond the
prediction of all the existing band structure calculations.
Particularly, the momentum and band dependence of the splitting, and
the anomalously small Stoner ratio (the ratio of exchange splitting
over magnetic moment) illustrate the unusual properties of the SDW
order. The detailed behaviors of the exchange splitting thus need to
be uncovered to further understand its microscopic origin.

One relevant question is whether SDW and superconductivity can
coexist at certain region of the phase diagram. Early resistivity
data have indirectly suggested that SDW and superconductivity could
coexist in LaO$_{1-\delta}$F$_{\delta}$FeAs\cite{JACS},
SmO$_{1-\delta}$F$_{\delta}$FeAs\cite{Liu}. However, more recent
neutron diffraction, muon spin relaxation ($\mu$SR), and
M\"{o}ssbauer spectroscopy indicate that they are exclusive from
each other for CeO$_{1-\delta}$F$_{\delta}$FeAs \cite{ZhaoNeutron}
and LaO$_{1-\delta}$F$_{\delta}$FeAs \cite{LuetkensNuclear}. The
anomaly in resistivity is associated with the structural transition
rather than the SDW.  On the other hand, the situation seems to be
quite different for Ba$_{1-x}$K$_x$Fe$_2$As$_2$, it has been shown
recently that the SDW and superconductivity could coexist  in
polycrystalline samples for $x\in(0.1,0.4)$ based on combined
transport, x-ray and neutron diffraction studies\cite{ChenBao}. If
one could rule out the caveat of possible phase segregation, this
would allude to a new ground state in Ba$_{1-x}$K$_x$Fe$_2$As$_2$,
where Cooper pairs are formed on a SDW background. This resembles
the Hg-based five-layer cuprate, where antiferromagnetic order
coexists with the superconductivity uniformly within single $CuO_2$
plane \cite{NMR}.  Novel properties might be expected.

\begin{figure}[b!]
\includegraphics[width=7cm]{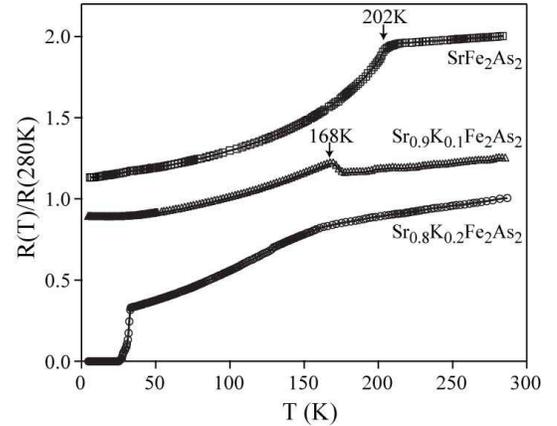}
\caption{Relative resistance (with respect to the resistance at
280K) of Sr$_{1-x}$K$_{x}$Fe$_2$As$_2$ ($x=0,0.1,0.2$) vs.
temperature. The $x=0$ and $x=0.1$ curves are shifted up by 0.25 and
1 respectively.} \label{Res}
\end{figure}

\begin{figure}[t]
\includegraphics[width=8cm]{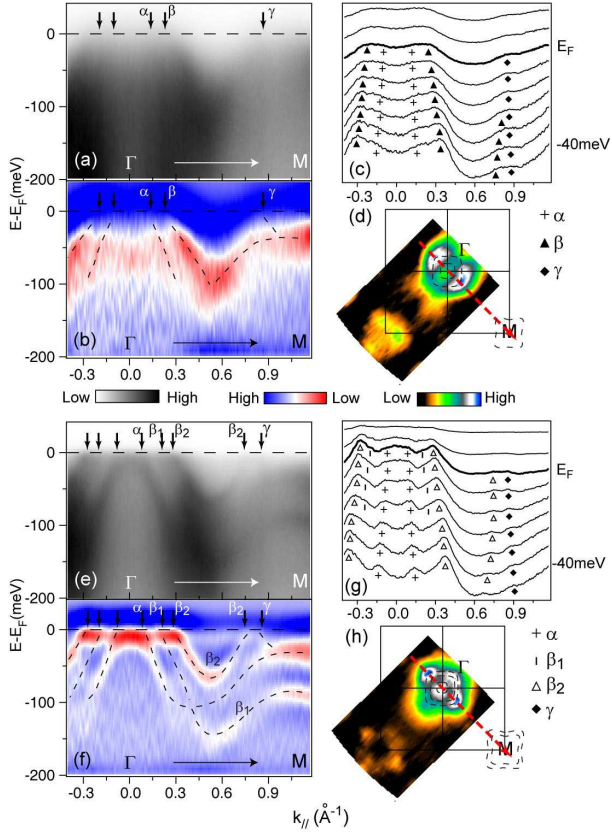}
\caption{(color online) Electronic structure of SrFe$_2$As$_2$. (a)
Photoemission intensity along the $\Gamma-M$ cut as indicated in
panel d. (b) The second derivative of the data in panel a. (c) The
MDC's near Fermi energy for the data in panel a. (d) Photoemission
intensity map at $E_F$ in the Brillouin zone, where the measured
Fermi surface sheets are shown by dashed curves. Only one set of
Fermi surface around M is shown for a clearer view. Data were taken
at 230K. (e,f,g,h) are the same as in panel a,b,c,d respectively,
but taken at 10K.}\label{Sr0}
\end{figure}

\begin{figure}[t]
\includegraphics[width=8cm]{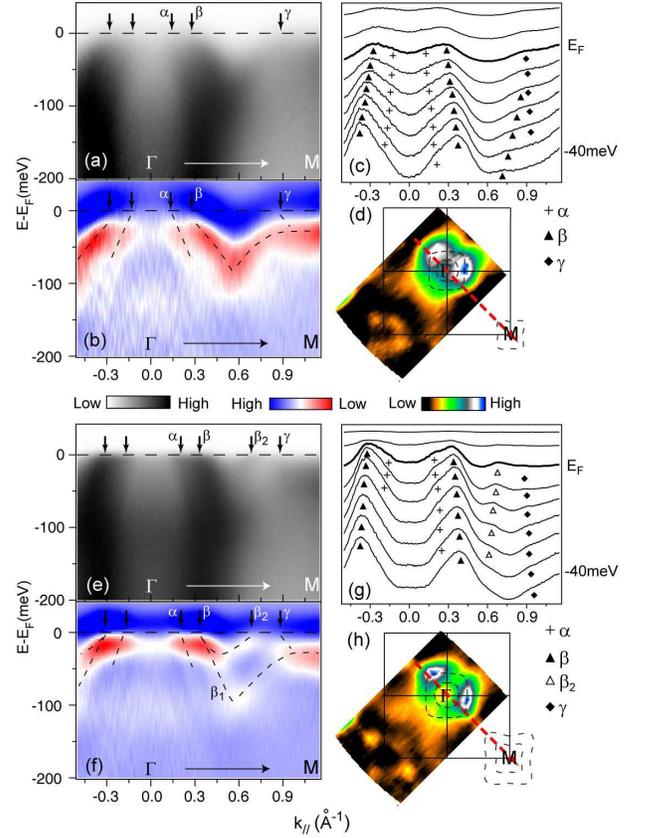}
\caption{(color online) Electronic structure of
Sr$_{0.8}$K$_{0.2}$Fe$_2$As$_2$.  (a) Photoemission intensity along
the $\Gamma-M$ cut as indicated in panel d. (b) The second
derivative of the data in panel a. (c) The MDC's near $E_F$ for the
data in panel a. (d) Photoemission intensity map at $E_F$ in the
Brillouin zone.  Data were taken at 150K. (e,f,g,h) are the same as
in panel a,b,c,d respectively, but taken at 10K.}\label{Sr3}
\end{figure}

In this Letter, we report angle resolved photoemission spectroscopy
(ARPES) measurements of Sr$_{1-x}$K$_x$Fe$_2$As$_2$ single crystals.
SrFe$_2$As$_2$ has the highest known SDW transition temperature
($T_S$) of about 205K in iron pnictides \cite{Sr}. We show that the
exchange splitting occurs in Sr$_{1-x}$K$_x$Fe$_2$As$_2$ for the
doping concentration $x=0,\,0.1,\,0.2$ with onset temperatures and
amplitudes in descending order. The systematics shows that the
exchange splitting is a fingerprint of the SDW on the electronic
structure. Therefore, our results on single crystalline samples
prove that superconductivity and SDW could coexist in
Sr$_{0.8}$K$_{0.2}$Fe$_2$As$_2$ (superconducting transition
temperature $T_c=25\,K$). The phase diagram of the
Sr$_{1-x}$K$_x$Fe$_2$As$_2$ thus would be very different from that
of the iron oxypnictide. Moreover, the quite different
manifestations of the exchange splitting in various systems further
highlight its complexity and correlated nature,  providing a new set
of clues for sorting out the microscopic mechanism of the splitting.

The Sr$_{1-x}$K$_x$Fe$_2$As$_2$ ($x=0,\,0.1,\,0.2$) single crystals
were synthesized with tin flux method \cite{singleSr}, where the
doping $x$ is determined through energy-dispersive x-ray (EDX)
analysis. The resistivity data in Fig.\,\ref{Res} indicate that the
undoped compound ($x=0$) enters the SDW state at about $202K$, and
there is an anomaly at 168K for $x=0.1$, while the $x=0.2$ compound
enters the zero resistance superconducting phase at about 25K. ARPES
measurements were performed with 24 eV photons from beamline 5-4 of
Stanford synchrotron radiation laboratory (SSRL) and beamline 9 of
Hiroshima synchrotron radiation center. With Scienta R4000 electron
analyzers, the overall energy resolution is 10meV, and angular
resolution is 0.3 degree. The samples were cleaved \textit{in situ},
and measured under ultra-high-vacuum of
$3\times10^{-11}$\textit{torr}.

\begin{figure*}[t!]
\includegraphics[width=17.5cm]{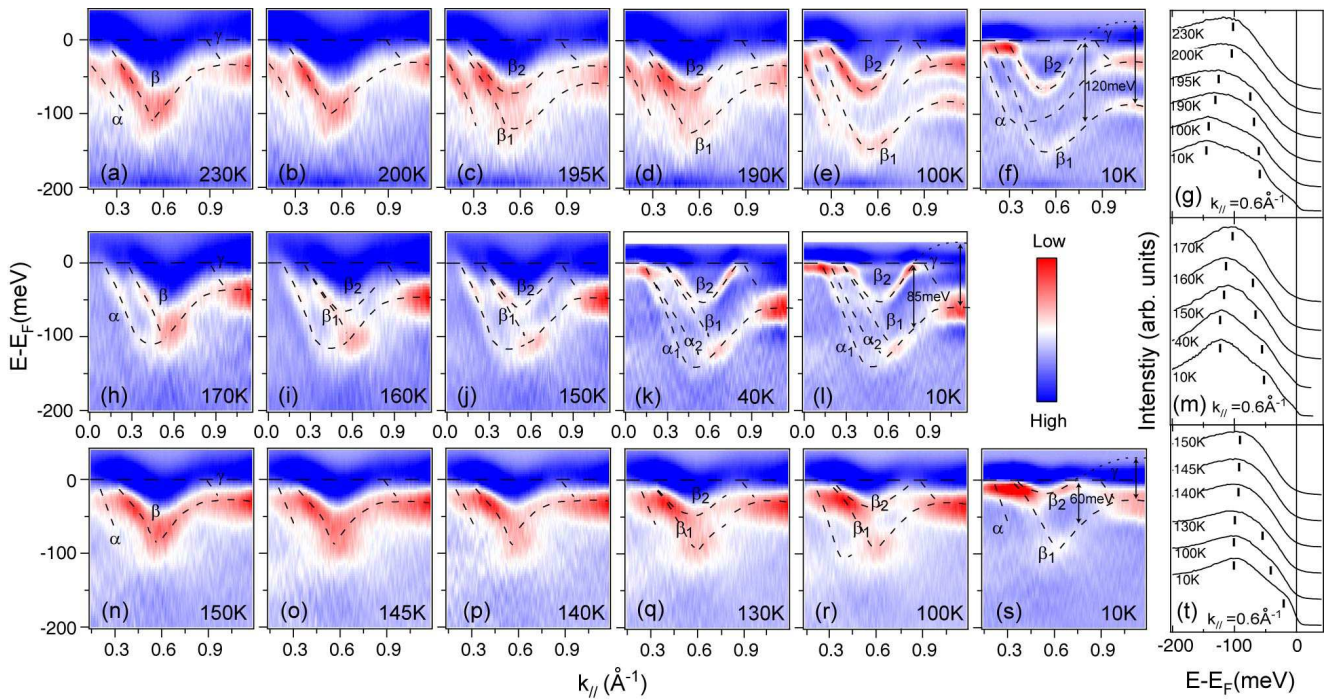}
\caption{(color online) Temperature dependence of the band
dispersion along the $\Gamma-M$ cut for Sr$_{1-x}$K$_x$Fe$_2$As$_2$.
Second derivative of photoemission intensity with respect to energy
(a-f) for $x=0$ at 230K, 200K, 195K, 190K, 100K, and 10K
respectively, (h-l) for $x=0.1$ at 170K, 160K, 150K, 40K, and 10K
respectively, and (n-s) for $x=0.2$ at 150K, 145K, 140K, 130K, 100K,
and 10K respectively. Dashed lines are the guides of eye for the
bands. Note the minimum of the second derivative represents a peak,
thus the lower part (red or white color) represents the band. (g),
(m) and (t) are the temperature evolution of EDC's at
$k=0.6\AA^{-1}$ for $x=0,0.1$ and $0.2$ respectively. Note the
momentum window is slightly wider for $x=0.1$ data.}\label{Tdep}

\end{figure*}

The normal state band structure of SrFe$_2$As$_2$ is presented
through the photoemission intensity and its second derivative with
respective to energy along the $\Gamma-M$ cut [Figs.\,\ref{Sr0}(a)
and \ref{Sr0}(b)]. Three bands (named as $\alpha$, $\beta$ and
$\gamma$ band respectively) could be identified to cross $E_F$, with
the assistance from the momentum distribution curves (MDC's) in
Fig.\,\ref{Sr0}(c). Near M, the $\alpha$ and $\beta$ bands become
quite flat and degenerate within the experimental resolution, and do
not cross the Fermi energy. There are thus two hole-type Fermi
surfaces around $\Gamma$, and one electron-type Fermi surface around
M [Fig.\,\ref{Sr0}(d)], as predicted by the band structure
calculations \cite{LDA0,LDA1,LDA2,LDA3,LDA4,LDA5,NLWang2}. In the
SDW state, the data along the same cut are measured for comparison
[Figs.\,\ref{Sr0}(e-g)]. Three Fermi crossings ($k_F$'s) could be
clearly resolved near $\Gamma$. The separation between the two
$\alpha$ $k_F$'s on both sides of $\Gamma$ is closer, giving a
smaller hole pocket than the normal state one. The $\beta$ band is
pushed away from $\Gamma$, and splits into two bands, which are
assigned as $\beta_{1}$ and $\beta_{2}$ respectively. Around M, the
normal-state flat feature splits into three bands and well connected
to features around $\Gamma$. Correspondingly, the $\beta_{1}$ band
is pushed down by about 60\,meV; the $\beta_{2}$ band is pushed up
to cross the Fermi energy; and the $\alpha$ band is more or less
unaffected. Moreover, the electron-like nature of the $\gamma$
pocket could be better resolved in Figs.\,\ref{Sr0}(f-g) than in the
normal state, and its $k_F$ does not show any noticeable movement.
Since Fermi surface folding in the SDW state is not observed, the
SDW state has two more hole pockets, one around $\Gamma$ and one
around M [Fig.\,\ref{Sr0}(h)] than the normal state. Similar to the
BaFe$_2$As$_2$ case, no energy gap is observed for all the bands at
their $k_F$'s, ruling out the ``Fermi-surface-nesting" mechanism for
SDW in itinerant electron systems like Chromium and its alloys
\cite{chromium}.

The corresponding electronic structure in the hole-doped
Sr$_{0.8}$K$_{0.2}$Fe$_2$As$_2$ superconductor is illustrated in
Fig.\,\ref{Sr3}. At high temperatures [Fig.\,\ref{Sr3}(a-d)], it is
similar to that in the normal state of SrFe$_2$As$_2$. As expected,
the two hole pockets around $\Gamma$ grow larger, and the electron
pocket around M slightly shrinks with hole doping. At 10K, there is
no obvious splitting near $\Gamma$. The most prominent difference
occurs midway in the $\Gamma$-M cut, where two features are observed
in Figs.\,\ref{Sr3}(f) and \ref{Sr3}(g), one of which (the $\beta_2$
band) crosses $E_F$, and gives an additional large hole pocket
around M at 10K in Fig.\,\ref{Sr3}(h).

To further illustrate the nature of splitting, detailed temperature
dependence of the bands in Sr$_{1-x}$K$_{x}$Fe$_2$As$_2$
($x=0,\,0.1,\,0.2$) are shown through the second derivative of the
photoemission intensity in Fig.\,\ref{Tdep}. For SrFe$_2$As$_2$,
although no obvious temperature dependence is observed for the
$\alpha$ band within the experimental resolution, the splitting of
the $\beta$ band occurs abruptly between 200K and 195K
Figs.\,\ref{Tdep}(b-c), and develops rapidly with the decreasing
temperatures. At the lowest temperature, the hybridization of the
$\alpha$ and $\beta_{1}$ could also be resolved clearly when they
cross. However, the bands are named as if they were not crossing.

For $x=0.1$  and $x=0.2$, band splittings occur very abruptly as
well. The onset temperatures are estimated to be $165\pm5K$ and
$135\pm5K$ for $x=0.1$ and $x=0.2$ respectively, as shown in
Figs.\,\ref{Tdep}(h-l) and Figs.\,\ref{Tdep}(n-s). The splitting is
momentum dependent in all cases. By extracting the largest splitting
between the $\beta_1$ and $\beta_2$ bands at the $k_F$ of $\beta_2$
(which are close to their splittings at M by fit), one gets
120\,meV, 85\,meV, and 60\,meV respectively for $x=0, 0.1$, and
$0.2$ respectively,  consistent with the decreasing onset
temperatures of the splitting.  As a comparison, the splitting
around $\Gamma$ is just about 50\,meV for $x=0$. We note for
BaFe$_2$As$_2$, $T_S=138K$, and the maximal splitting is about 75\,
meV near M \cite{YangExchange}; both are close to the
Sr$_{0.9}$K$_{0.1}$Fe$_2$As$_2$ case. Furthermore, all systems show
similar spectral characters when the splitting are the most obvious.
For example, the temperature evolutions of photoemission spectra at
$k=0.6\AA^{-1}$ are quite similar in Figs.\,\ref{Tdep}(g),
\ref{Tdep}(m) and \ref{Tdep}(t) for $x=0,\,0.1,\,0.2$ respectively.

The band splitting occurs almost exactly at their bulk SDW
transition temperatures for both SrFe$_2$As$_2$ and BaFe$_2$As$_2$,
and at the resistivity anomaly temperature of
Sr$_{0.9}$K$_{0.1}$Fe$_2$As$_2$. Considering that $T_S$ drops
rapidly with doping \cite{ChenBao}, plus the drastically different
low temperature band structures, one can conclude that the measured
electronic structure reflects the bulk properties, and rule out any
phase separation effects in all data. Similar to BaFe$_2$As$_2$
\cite{YangExchange}, such a splitting on the order of several $k_B
T_S$ and its temperature dependence cannot be explained by factors
such as structure transition or spin orbital coupling. Instead, it
can be most naturally explained by the exchange splitting associated
with the SDW formation. In fact, the electronic energy of the system
can be saved through such a splitting, and thus it can be
responsible for the SDW. Consistently, the band splitting is of the
same scale as the exchange interactions between the nearest and
next-nearest neighbor local moments estimated from LDA
calculations\cite{YangExchange,Lu}. In this regard, the observed
systematics, such as the correlations among doping/onset
temperature/splitting amplitude, and similar spectral characters
indicate that the origin of band splittings in
Sr$_{0.8}$K$_{0.2}$Fe$_2$As$_2$ is no different from others.
Therefore, our results on single crystalline samples provide a
compelling piece of evidence for the coexistence of the SDW and
superconductivity in an iron pnictide.


The paramagnetic state electronic structures of various iron
pnictides qualitatively resemble each other
\cite{YangExchange,DHLu,XJZhouSC,DHSC,Kami}, regardless of the
chemical environment or doping, as exemplified here for
Sr$_{1-x}$K$_{x}$Fe$_2$As$_2$. Nevertheless, Fig.\ref{Tdep} also
illustrates that the detailed behaviors of exchange splitting  in
various system can be rather different besides their similarities
mentioned above. Take the splitting at M as an example, the shifts
of both the $\beta_1$ and $\beta_2$ bands are equally strong from
the normal state position, and the $\alpha$ band does not split for
$x=0$; for $x=0.1$, $\beta_2$ shifts much more than $\beta_1$, and
its $\alpha$ band shows a shift; for $x=0.2$, only $\beta_2$ shows
obvious shift. While for BaFe$_2$As$_2$, all bands shift strongly at
M\cite{YangExchange}. Particularly, the electron Fermi pocket around
M splits into one large and one small electron pockets in
BaFe$_2$As$_2$; but for SrFe$_2$As$_2$, the size of the $\gamma$
pocket does not change noticeably, indicating a negligible
splitting. Similarly, one could find further differences for
exchange splitting around $\Gamma$ as well. These nontrivial
findings unveil the correlated/non-rigid-band aspect of the exchange
splitting.


The coexistence of SDW and superconductivity has profound
consequences on the nature of the superconductivity.  It not only
suggests that the  superconducting gap might open at one more
($\beta_2$) Fermi surface sheet in this material than in the
Ba$_{0.6}$K$_{0.4}$Fe$_2$As$_2$  reported earlier
\cite{XJZhouSC,DHSC}. Because a split band is either majority or
minority band that is in-phase or out-of-phase with the SDW spin
order respectively, take a Cooper pair based on electrons at $\pm
k_F$ of a majority band in the singlet pairing channel for example,
its spin-up electron and spin-down electron must be mainly situated
in the spin-up and spin-down sites of the SDW respectively. This
gives a novel ground state that is not known before. Moreover,
because the SDW does not open an gap at $k_F$, how it competes with
the superconductivity in iron pnictides would be another interesting
issue. On the other hand, the magnetic fluctuations related to SDW
might even play a constructive role in superconductivity as in
cuprates. We leave the detailed studies of these issues to future.


To summarize, we have measured the electronic structures of
Sr$_{1-x}$K$_{x}$Fe$_2$As$_2$. We show that besides the quantitative
differences, the detailed behaviors of the splitting differ
prominently in various iron pnictides. Since band structure
calculations so far failed to reproduce or predict the observed
exchange splitting, our results provide important new clues for
revealing the microscopic origin of the exchange splitting and SDW
in iron pnictides. Particularly, we show in the single crystalline
Sr$_{0.8}$K$_{0.2}$Fe$_2$As$_2$ that SDW and superconductivity could
coexist, revealing a new kind of ground state, which would  help
understand the relationship between the SDW and superconductivity in
iron pnictides.

We thank Dr. W. Bao for inspiring discussions, and Dr. D. H. Lu and
Dr. R. G. Moore for their experimental assistance at SSRL. This work
was supported by the Nature Science Foundation of China, the
Ministry of Science and Technology of China (National Basic Research
Program No.2006CB921300), and STCSM. SSRL is operated by the DOE
Office of Basic Energy Science under Contract No. DE-AC03-765F00515.

Note added in support: During preparation of this manuscript, a
similar finding on SDW/superconductivity coexistence in
(Ba,K)Fe$_2$As$_2$, (Sr,Na)Fe$_2$As$_2$, and CaFe$_2$As$_2$ is
posted online based on $\mu$SR experiments\cite{muSR}.


\end{document}